\documentclass[%
notitlepage,%
onecolumn,%
floats,%
aps,%
prd,%
nobibnotes,%
nofootinbib,%
amsmath,%
amssymb,%
amsfonts,%
amscd,%
superscriptaddress,%
preprint
]{revtex4-1}
\usepackage[utf8]{inputenc}
\usepackage[left=20mm,right=20mm,top=20mm,bottom=25mm]{geometry}
\usepackage[caption=false]{subfig}
\usepackage{tikz}
\usepackage{hyperref}
\usepackage{array}

\newcommand{\Br}{\mathrm{Br}}

\usetikzlibrary{trees,arrows}
\usetikzlibrary{decorations.pathmorphing}
\usetikzlibrary{decorations.markings}

\tikzset{
  >=stealth',
 every node/.style={font=\scriptsize},
    photon/.style={decorate, decoration={snake,amplitude=2pt,segment
        length=8pt,pre length=0cm,post length=0}},
    photontiny/.style={decorate, decoration={snake,amplitude=1pt,segment
        length=4pt,pre length=0cm,post length=0}},
    electron/.style={postaction={decorate},
        decoration={markings,mark=at position .55 with {\arrow{>}}}},
    antielectron/.style={postaction={decorate},
        decoration={markings,mark=at position .52 with {\arrow{<}}}},
    drelectron/.style={line width=1.5pt,postaction={decorate},
        decoration={markings,mark=at position .52 with {\arrow{>}}}},
    gluon/.style={decorate,
        decoration={coil,amplitude=4pt, segment length=5pt}},
    scalar/.style={densely dashed}
}
\begin{document}

\title{Dimuon resonance near $28$~GeV and muon anomaly}

\author{S.~I.~Godunov}
\email{sgodunov@itep.ru}
\affiliation{\small Institute for Theoretical and Experimental Physics,
117218, Moscow, Russia}
\affiliation{\small Novosibirsk State University,
630090, Novosibirsk, Russia}

\author{V.~A.~Novikov}
\email{novikov@itep.ru}
\affiliation{\small Institute for Theoretical and Experimental Physics,
117218, Moscow, Russia}
\affiliation{\small Moscow Institute of Physics and Technology, 141700,
Dolgoprudny, Moscow Region, Russia}
\affiliation{\small National Research University Higher School of
  Economics, 101978, Moscow, Russia}

\author{M.~I.~Vysotsky}
\email{vysotsky@itep.ru}
\affiliation{\small Institute for Theoretical and Experimental Physics,
117218, Moscow, Russia}
\affiliation{\small Moscow Institute of Physics and Technology, 141700,
Dolgoprudny, Moscow Region, Russia}
\affiliation{\small National Research University Higher School of
  Economics, 101978, Moscow, Russia}

\author{E.~V.~Zhemchugov}
\email{zhemchugov@itep.ru}
\affiliation{\small Institute for Theoretical and Experimental Physics,
117218, Moscow, Russia}
\affiliation{\small National Research Nuclear University Moscow Engineering Physics Institute,
115409, Moscow, Russia}

\begin{abstract}
  We discuss if the resonance recently observed by CMS can be
  responsible for the deviation of the experimentally measured muon
  anomalous magnetic moment from the theoretical prediction.
\end{abstract}

\maketitle

\newpage

\section{Introduction}

The CMS collaboration has recently reported a peak at invariant mass
\begin{equation}
  m_X = 28.3 \pm 0.4~\text{GeV}
\end{equation}
of $\mu^+ \mu^-$ pairs produced in association with $b$ jet in
$pp$-collisions at the LHC~\cite{cms}. The peak appeared in the 8~TeV
data with $19.7~\text{fb}^{-1}$ of integrated luminosity, while no
significant excess was found in the 13~TeV data with
$35.9~\text{fb}^{-1}$ of integrated luminosity.  The observation was
made for two event categories with different cuts on jets directions
with the local significancies of $4.2$ and $2.9$ standard deviations
(see the paper for the details). The fiducial cross section for both
categories is at the level of 4~fb. Signal selection efficiency can
strongly depend on the production process, so to evaluate the total
$\sigma \times \Br(X \to \mu^+ \mu^-)$ a particular model is
required. The CMS paper does not study any specific model, so only the
fiducial cross sections were provided.

The reported width of the peak is
\begin{equation}
  \Gamma^\text{(exp.)}_X = 1.8 \pm 0.8~\text{GeV}
  \label{X-width}
\end{equation}
which is several times larger than the expected mass resolution for a
dimuon system $\sigma_{\mu \mu} = 0.45$~GeV.

We shall study whether the resonance $X$ (if its existence will be
confirmed in the future) can explain the deviation of the measured
value of the muon anomalous magnetic moment
$a_\mu \equiv (g - 2)_\mu / 2$ from the Standard Model value
\begin{equation}
  \delta a_\mu
  \equiv a_\mu^\text{exp.} - a_\mu^\text{SM}
  = \left\{
     \begin{aligned}
       (31.3 &\pm 7.7) \cdot 10^{-10},\text{ see \cite{1705.00263}}, \\
       (26.8 &\pm 7.6) \cdot 10^{-10},\text{ see \cite{1706.09436}}.
     \end{aligned}
   \right.
\end{equation}
In the following numerical estimates we will use the average of these
two values:
\begin{equation}
  \delta a_\mu = (29 \pm 8) \cdot 10^{-10}.
  \label{amu-experimental}
\end{equation}

\section{$X$ contributions to $\delta a_\mu$}
\label{sec:g-2}

\begin{figure}
  \centering
  \subfloat[scalar and pseudoscalar~$X$]{
    \begin{tikzpicture}[]
      %    A
      %    |
      %    B
      %    /\
      %   C--D
      %  /    \
      % E      F
      \coordinate (A) at (0,0);
      \coordinate (B) at (0,-2);
      \coordinate (C) at (-1.4,-2.7);
      \coordinate (D) at (1.4,-2.7);
      \coordinate (E) at (-2,-3);
      \coordinate (F) at (2,-3);
      \draw[photon] node[right]{$\gamma$} (A) to (B);
      \draw[electron] (C) to (B);
      \draw[electron] (B) to (D);
      \draw[scalar]   (C) to node[below]{$S$, $P$} (D);
      \draw[electron] (E) to node[below]{$\mu$} (C);
      \draw[electron] (D) to node[below]{$\mu$} (F);
      \filldraw (B) circle (1.5pt);
      \filldraw (C) circle (1.5pt);
      \filldraw (D) circle (1.5pt);
     \end{tikzpicture}
  }
  \hspace{20pt}
  \subfloat[vector and axial vector~$X$]{
    \begin{tikzpicture}[]
      %    A
      %    |
      %    B
      %    /\
      %   C--D
      %  /    \
      % E      F
      \coordinate (A) at (0,0);
      \coordinate (B) at (0,-2);
      \coordinate (C) at (-1.4,-2.7);
      \coordinate (D) at (1.4,-2.7);
      \coordinate (E) at (-2,-3);
      \coordinate (F) at (2,-3);
      \draw[photon] node[right]{$\gamma$} (A) to (B);
      \draw[electron] (C) to (B);
      \draw[electron] (B) to (D);
      \draw[photon]   (C) to node[below]{$V$, $A$} (D);
      \draw[electron] (E) to node[below]{$\mu$} (C);
      \draw[electron] (D) to node[below]{$\mu$} (F);
      \filldraw (B) circle (1.5pt);
      \filldraw (C) circle (1.5pt);
      \filldraw (D) circle (1.5pt);
     \end{tikzpicture}
   }
   \caption{One-loop contribution of $X$ to $\delta a_\mu$.}
   \label{f:amu-1-loop}
\end{figure}
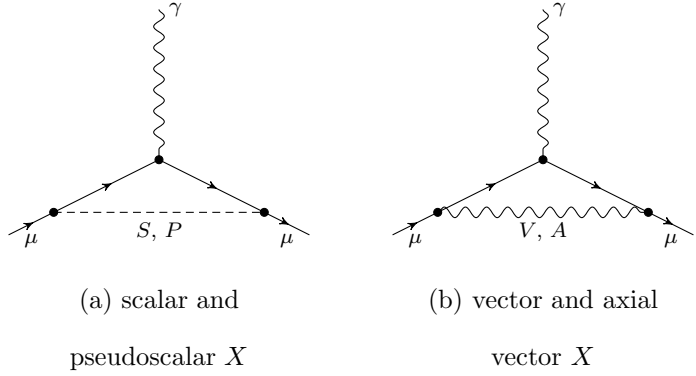

Let us consider the Standard Model extended with a field $X$.  Its
contribution to the muon anomalous magnetic moment depends on $X$
spin. We will consider the following four possibilities: scalar $S$,
pseudoscalar $P$, vector $V$, axial vector $A$. Their coupling to
muons is described by the following terms in the Lagrangian:
\begin{equation}
  \begin{aligned}
    \Delta \mathcal{L}_{S \mu \mu}
    &= Y_{S \mu \mu} \, \bar \mu \mu \, S
    && \text{(scalar $X$)}, \\
    \Delta \mathcal{L}_{P \mu \mu}
    &= i Y_{P \mu \mu} \, \bar \mu \gamma_5 \mu \, P
    && \text{(pseudoscalar $X$)}, \\
    \Delta \mathcal{L}_{V \mu \mu}
    &= Y_{V \mu \mu} \, \bar \mu \gamma_\mu \mu \, V_\mu
    && \text{(vector $X$)}, \\
    \Delta \mathcal{L}_{A \mu \mu}
    &= Y_{A \mu \mu} \, \bar \mu \gamma_\mu \gamma_5 \mu \, A_\mu
    && \text{(axial vector $X$)}.
  \end{aligned}
  \label{mu-coupling}
\end{equation}
An exchange of $X$ contributes at one loop to $a_\mu$ (see
Fig.~\ref{f:amu-1-loop}).
The following results were obtained in~\cite[Eq. (260)]{0902.3360}:
\begin{align}
  \delta a_\mu^S
  &= \frac{Y_{S \mu \mu}^2}{4 \pi^2}
     \left( \frac{m_\mu}{m_X} \right)^2
     \left[ \ln \frac{m_X}{m_\mu} - \frac{7}{12} \right]
  && \text{(scalar $X$)},
  \\
  \delta a_\mu^P
  &= \frac{Y_{P \mu \mu}^2}{4 \pi^2}
     \left( \frac{m_\mu}{m_X} \right)^2
     \left[ -\ln \frac{m_X}{m_\mu} + \frac{11}{12} \right]
  && \text{(pseudoscalar $X$)},
  \\
  \delta a_\mu^V
  &= \frac{Y_{V \mu \mu}^2}{4 \pi^2}
     \left( \frac{m_\mu}{m_X} \right)^2
     \cdot \frac13
  && \text{(vector $X$)},
  \\
  \delta a_\mu^A
  &= \frac{Y_{A \mu \mu}^2}{4 \pi^2}
     \left( \frac{m_\mu}{m_X} \right)^2
     \cdot \left( -\frac53 \right)
  && \text{(axial vector $X$)},
\end{align}
where $m_X \gg m_\mu$ is supposed. Only the scalar and vector $X$ can
resolve the
discrepancy~\eqref{amu-experimental}. Equating~\eqref{amu-experimental} 
to $\delta a_\mu^S$ and $\delta a_\mu^V$ results in
\begin{equation}
  \begin{aligned}
    Y_{S \mu \mu} &= 0.041 \pm 0.006, \\
    Y_{V \mu \mu} &= 0.16 \pm 0.02.
  \end{aligned}
  \label{Y-mu-mu}
\end{equation}
In this case the $X \to \mu^+ \mu^-$ decay width
\begin{equation}
  \begin{aligned}
    \Gamma(S \to \mu^+ \mu^-)
    &= \frac{Y_{S \mu \mu}^2}{8 \pi}
       m_X
       \left(1 - \frac{4 m_\mu^2}{m_X^2} \right)^{3/2}
    &&= 1.8 \pm 0.5~\text{MeV}, \\
    \Gamma(V \to \mu^+ \mu^-)
    &= \frac{Y_{V \mu \mu}^2}{8 \pi}
       m_X
       \sqrt{1 - \frac{4 m_\mu^2}{m_X^2}}
    &&= 28 \pm 8~\text{MeV},
  \end{aligned}
\end{equation}
and the corresponding branching ratios
\begin{equation}
  \Br(X \to \mu^+ \mu^-)
  = \frac{\Gamma(X \to \mu^+ \mu^-)}{\Gamma_X^\text{(exp.)}}
  = \left\{
      \begin{aligned}
        &(1.0 \pm 0.5) \cdot 10^{-3} && \text{ for } S \to \mu^+ \mu^-, \\
        &(1.5 \pm 0.8) \cdot 10^{-2} && \text{ for } V \to \mu^+ \mu^-.
      \end{aligned}
    \right.
\end{equation}

Since the uncertainty in the measurement of $\Gamma_{X}$ is rather
large, the $X \to \mu^+ \mu^-$ decay can dominate or even be the only
decay~of~$X$.

Another possibility is that $X$ can decay to other particles.  For the
scalar, such a small branching ratio can be naturally explained if $S$
couples to $\tau^+ \tau^-$ as well, and the coupling constants are
proportional to $\mu$ and $\tau$ masses correspondingly. Then
\begin{equation}
  \Gamma(S \to \tau^+ \tau^-) =
  \left(\frac{m_{\tau}}{m_{\mu}}\right)^{2}
  \Gamma\left(S\to\mu^{+}\mu^{-}\right) = 0.52 \pm 0.15~\text{GeV},
  \label{X->tau-tau-branching}
\end{equation}
which is in agreement with the reported value~\eqref{X-width}.

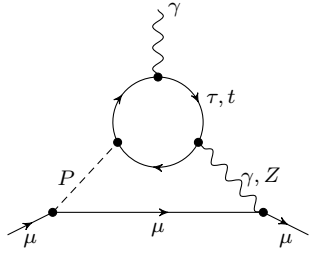
\begin{figure}[t]
  \centering
  \begin{tikzpicture}[]
    %      A
    %      |
    %      B
    %     /  \
    %    /    \
    %    \    /
    %     C--D
    %    /    \
    %   E------F
    %  /        \
    % G          H
    \coordinate (A) at (0,0);
    \coordinate (B) at (0,-0.9);
    \coordinate (C) at (-0.5366563145999494,-1.7683281572999747);
    \coordinate (D) at (0.5366563145999494,-1.7683281572999747);
    \coordinate (E) at (-1.4,-2.7);
    \coordinate (F) at (1.4,-2.7);
    \coordinate (G) at (-2.0,-3);
    \coordinate (H) at (2.0,-3);
    
    \draw[photon] node[right]{$\gamma$} (A) to (B);
    \draw[electron] (B) arc (90:-26.56505117707799:0.6) node[midway,right]{$\tau,t$};
    \draw[electron] (C) arc (206.56505117707798:90:0.6);
    \draw[electron] (D) arc (-26.56505117707799:-153.43494882292202:0.6);
    \draw[electron] (G) to node[below]{$\mu$} (E);
    \draw[electron] (E) to node[below]{$\mu$} (F);
    \draw[electron] (F) to node[below]{$\mu$} (H);
    \draw[scalar]   (E) to node[left]{$P$} (C);
    \draw[photon]   (F) to node[right]{$\gamma,Z$} (D);
    \filldraw (B) circle (1.5pt);
    \filldraw (C) circle (1.5pt);
    \filldraw (D) circle (1.5pt);
    \filldraw (E) circle (1.5pt);
    \filldraw (F) circle (1.5pt);
   \end{tikzpicture}
  \caption{Two-loop contribution of $P$ to $\delta a_\mu$.}
  \label{f:amu-2-loop}
\end{figure}

One of the most natural generalizations of the SM is the model with
additional heavy Higgs doublet, the so-called two Higgs doublets model
(2HDM). Quite unexpectedly, the leading contributions to $a_\mu$ in
this model for some values of parameters arise at the two-loop level
(see Fig.~\ref{f:amu-2-loop}), and light spin zero particle is needed
to compensate the two-loop suppression~\cite{hep-ph/0009292,
  1409.3199, 1605.06298, 1412.4874, 1504.07059, 1507.07567,
  1507.08067, 1511.05162, 1502.04199, 1607.06292}. It was found that a
light pseudoscalar boson $P$ with strong couplings to leptons could
explain the current value of $\delta
a_\mu$~\eqref{amu-experimental}. According to a recent
paper~\cite{1711.11567}, in a very small parameter region around
$m_A = 20$~GeV the extra contribution to $a_\mu$ even exceeds the one
needed to explain deviation~\eqref{amu-experimental}. That is why it
looks very appealing to identify the resonance found in~\cite{cms} as
the pseudoscalar boson $P$ from 2HDM, resolving simultaneously the
problem with muon anomaly. For this reason we will not discard
pseudoscalar $P$ from consideration yet.

\section{LEP data and $X$}

If $X$ is responsible for the muon anomaly then we know $X$ coupling
to muons, see Section~\ref{sec:g-2}. In this section we are going to
investigate how $X$ modifies $Z$ boson properties.

The width of $Z$ decay to a fermion-antifermion pair and a
pseudoscalar is~\cite{plb259.175}
\begin{equation}
  \Gamma(Z \to f \bar f X)
  = \frac{\alpha}{128 \pi^2}
    \frac{m_Z}{\sin^2 \theta_W \cos^2 \theta_W}
    \frac{N_c}{3}
    Y_{X f \bar f}^2
    ((g_A^2 + g_V^2) F_1 + (g_V^2 - g_A^2) F_2),
  \label{djouadi}
\end{equation}
where $N_c$ is the number of fermion colors, $g_V$ and $g_A$ are the
axial and vector couplings of the fermion to the $Z$ boson
($g_V = T_3$, $g_A = T_3 - 2 Q \sin^2 \theta_W$, $T_3$ is the third
component of the weak isospin, and $Q$ is the electric charge of the
fermion),
\begin{equation}
  \begin{aligned}
    F_1 &= -2 (1 + 3a) \ln a + \tfrac13 (1 - a) (a^2 - 8a - 17), \\
    F_2 &= 2a (5 + 3a) \ln a
         - \tfrac13 (1 - a) (a^2 - 44a - 5)
    \\ & \quad
         + 4 a^2 \left[
            \tfrac12 \ln^2 a
            - \ln a \ln(1 + a)
            + \mathrm{Li}_2 \left( \frac{a}{1 + a} \right)
            - \mathrm{Li}_2 \left( \frac{1}{1 + a} \right)
           \right],
  \end{aligned}
\end{equation}
$a = m_X^2 / m_Z^2$, and $\mathrm{Li}_2(x)$ is the dilogarithm,
\begin{equation*}
  \mathrm{Li}_2(x) = -\int\limits_0^x \frac{\ln(1 - z)}{z} \, \mathrm{d}z.
\end{equation*}
In this formula the fermion is assumed to be massless, and in this
limit it also works for the scalar $X$.

The $X$ particle will provide an extra contribution to $Z \to 4 \mu$
decay through the following process:
$Z \to \mu^+ \mu^- X(\to \mu^+ \mu^-)$. According to~\eqref{djouadi},
\begin{equation}
  \Gamma(Z \to \mu^+ \mu^- X)
  = 6.4 \cdot 10^{-5} \times Y_{S \mu \mu}^2~\text{GeV}
  \approx 105~\text{eV},
\end{equation}
where the value of $Y_{S\mu\mu}$ from~\eqref{Y-mu-mu} was
substituted. Hence
\begin{equation}
  \Br(Z \to \mu^+ \mu^- X(\to \mu^+ \mu^-))
  \approx 4.2 \cdot 10^{-8} \, \Br(X \to \mu^+ \mu^-),
\end{equation}
and even for $\Br(X \to \mu^+ \mu^-) = 1$ it is one order of magnitude
less than the experimental error:
$\Br(Z \to 4 \ell) = (3.5 \pm 0.4) \cdot 10^{-6}$~\cite{pdg}.

The width of $X$ of the order of 1~GeV may be explained by
$X \to \tau^+ \tau^-$ and\slash{}or $X \to \nu \bar \nu$ decays. The
upper limit on the $Y_{X \tau \tau}$ coupling can be obtained from the
results of the DELPHI collaboration on the search of
$Z \to \tau^+ \tau^- h (\to \tau^+ \tau^-)$ decays. According
to~\cite[Fig.~11]{hep-ex/0410017}, the value of
$Y_{X \tau \tau} = 100 \, m_\tau / v \approx 0.7$ is allowed at 95\%
C.L., where $v \approx 246$~GeV is the Higgs boson expectation
value. In this case $\Gamma_X \approx 0.6$~GeV both for the scalar and
pseudoscalar $X$, which is in agreement with the
estimate~\eqref{X->tau-tau-branching}.

$X \to \nu \bar \nu$ decay increases the invisible $Z$ boson width by
the following quantity:
\begin{equation}
  \Gamma(Z \to \nu \bar \nu X) = 0.4 \, Y_{X \nu \nu}^2~\text{MeV}.
\end{equation}
Since experimental uncertainty in the value of
$\Gamma(Z \to \text{invisible})$ is about $1.5$~MeV, the value of
$Y_{X \nu \bar \nu}$ of the order of one is allowed leading to a GeV
width of $X \to \nu \bar \nu$ decay.

\section{Can $X$ be produced via radiation from $b$ quark?}

The $X$ boson is seen by the CMS in association with at least one
$b$-tagged jet. Let us consider if it can be produced via radiation
from $b$ quark.  Let the coupling of $X$ with $b$-quarks be described
by interactions analogous to~\eqref{mu-coupling}:
\begin{equation}
  \begin{aligned}
    \Delta \mathcal{L}_{S b b}
    &= Y_{S b b} \, \bar b b \, S
    && \text{(scalar $X$)}, \\
    \Delta \mathcal{L}_{P b b}
    &= i Y_{P b b} \, \bar b \gamma_5 b \, P
    && \text{(pseudoscalar $X$)}, \\
    \Delta \mathcal{L}_{V b b}
    &= Y_{V b b} \, \bar b \gamma_\mu b \, V_\mu
    && \text{(vector $X$)}, \\
    \Delta \mathcal{L}_{A b b}
    &= Y_{A b b} \, \bar b \gamma_\mu \gamma_5 b \, A_\mu
    && \text{(axial vector $X$)}.
  \end{aligned}
  \label{b-coupling}
\end{equation}

In Ref.~\cite{cms}, the CMS collaboration reports fiducial cross
sections for two event categories. In both cases exactly two jets with
high $p_T$ are required, one of which is $b$-tagged, and the
$b$-tagged jet has to be in the barrel region. The main difference
between the categories is in the direction of the untagged jet: it can
be in either the endcap or the barrel regions. In the following the
first event category will be considered since it possesses the highest
significance of $4.2$~standard deviations. The corresponding fiducial
cross section is
\begin{equation}
  \sigma_\text{fid.1} = 4.1 \pm 1.4~\text{fb},
  \label{eq:fid.1}
\end{equation}
and the cuts are summarized in Table~1 from~\cite{cms}.

To calculate the cross section of $X$ production at the LHC, CalcHEP
3.6.30~\cite{calchep} was used. CalcHEP parameters were updated to
their modern values according to Ref.~\cite{pdg}. \linebreak
MMHT2014nnlo68cl~\cite{0901.0002} from the Les Houches PDF
library~\cite{1412.7420} was used as the set of parton distribution
functions.

Calculated cross sections for the first event category cuts (fiducial
cross sections) are presented in Table~\ref{t:pp->bXjet_cut}. Thus,
the events with two $b$ jets correspond to approximately one sixth of
the reported fiducial cross section~(\ref{eq:fid.1}).
\renewcommand{\extrarowheight}{1.7mm}
\newlength{\spl}
\setlength{\spl}{20mm}

\begin{table}[p]
  \centering
  \caption{ Fiducial cross sections $\sigma^{\rm fid}$ for the
    $pp \to b X + jet + \dots$ reaction and its subprocesses for
    $Y_{X b \bar b} = 0.01$ and $Y_{X\mu\mu}=1$ at $\sqrt{s} =
    8$~TeV. We took such a small value of $Y_{Xbb}$ to suppress
    multiple $X$ exchanges. The errors correspond to integration
    errors reported by CalcHEP. When summing up one should multiply
    the value by two if there are two reactions in left column. The
    second column corresponds to the multiplicity due to the two
    possibilities of the quark and its parent proton
    combination and due to the fact that each $b$ jet can be
    directed into barrel if there are more than one $b$ jet.}
  \label{t:pp->bXjet_cut}
  \begin{tabular}{|c|c|l|l|l|l|}
    \hline
    Subprocess & Mult.
    & $\sigma^{\rm fid}\cdot10^{5}$ [pb], $S$
    & $\sigma^{\rm fid}\cdot10^{5}$ [pb], $P$
    & $\sigma^{\rm fid}\cdot10^{5}$ [pb], $V$
    & $\sigma^{\rm fid}\cdot10^{5}$ [pb], $A$\\
    \hline
    \begin{minipage}{\spl}
      \centering
      $bu\to ub\mu\mu$\\
      $\bar bu\to u\bar b\mu\mu$
    \end{minipage}
               & 2
    & $5.23(2)$
    & $5.22(2)$
    & $16.0(1)$
    & $16.3(1)$\\
    \hline
    \begin{minipage}{\spl}
      \centering
      $b\bar u\to \bar ub\mu\mu$\\
      $\bar b\bar u\to \bar u\bar b\mu\mu$
    \end{minipage}
               & 2
    & $0.298(1)$
    & $0.293(2)$
    & $0.86(1)$
    & $0.89(1)$\\
    \hline
    \begin{minipage}{\spl}
      \centering
      $bd\to db\mu\mu$\\
      $\bar bd\to d\bar b\mu\mu$
    \end{minipage}
               & 2
    & $2.30(1)$
    & $2.30(1)$
    & $6.91(4)$
    & $7.14(4)$\\
    \hline
    \begin{minipage}{\spl}
      \centering
      $b\bar d\to \bar db\mu\mu$\\
      $\bar b\bar d\to \bar d\bar b\mu\mu$
    \end{minipage}
               & 2
    & $0.359(2)$
    & $0.355(2)$
    & $1.05(1)$
    & $1.08(1)$\\
    \hline
    \begin{minipage}{\spl}
      \centering
      $bs\to sb\mu\mu$\\
      $\bar bs\to s\bar b\mu\mu$
    \end{minipage}
               & 2
    & $0.209(2)$
    & $0.202(1)$
    & $0.593(4)$
    & $0.617(5)$\\
    \hline
    \begin{minipage}{\spl}
      \centering
      $b\bar s\to \bar sb\mu\mu$\\
      $\bar b\bar s\to \bar s\bar b\mu\mu$
    \end{minipage}
               & 2
    & $0.206(2)$
    & $0.205(1)$
    & $0.602(5)$
    & $0.618(6)$\\
    \hline
    \begin{minipage}{\spl}
      \centering
      $bc\to cb\mu\mu$\\
      $\bar bc\to c\bar b\mu\mu$
    \end{minipage}
               & 2
    & $0.113(1)$
    & $0.114(1)$
    & $0.336(2)$
    & $0.350(3)$\\
    \hline
    \begin{minipage}{\spl}
      \centering
      $b\bar c\to \bar cb\mu\mu$\\
      $\bar b\bar c\to \bar c\bar b\mu\mu$
    \end{minipage}
               & 2
    & $0.115(1)$
    & $0.114(1)$
    & $0.337(3)$
    & $0.340(2)$\\
    \hline
    \begin{minipage}{\spl}
      \centering
      $bg\to gb\mu\mu$\\
      $\bar bg\to g\bar b\mu\mu$
    \end{minipage}
               & 2
    & $7.42(8)$
    & $7.54(8)$
    & $22.1(2)$
    & $23.4(3)$\\
    \hline
    \begin{minipage}{\spl}
      \centering
      $bb\to bb\mu\mu$\\
      $\bar b\bar b\to \bar b\bar b\mu\mu$
    \end{minipage}
               & 1
    & $0.146(3)$
    & $0.142(2)$
    & $0.36(1)$
    & $0.45(1)$\\
    \hline
    $gg\to b\bar b\mu\mu$
               & 2
    & $5.19(8)$
    & $5.11(7)$
    & $21.3(3)$
    & $20.6(5)$\\
    \hline    
    \begin{minipage}{\spl}
      \centering
      $b\bar b\to b\bar b\mu\mu$
    \end{minipage}
               & 4
    & $0.082(1)$
    & $0.085(3)$
    & $0.286(3)$
    & $0.222(3)$\\
    \hline
    \begin{minipage}{\spl}
      \centering
      $u\bar u\to b\bar b\mu\mu$
    \end{minipage}
               & 4
    & $0.0636(3)$
    & $0.0631(3)$
    & $0.182(2)$
    & $0.184(2)$\\
    \hline
    \begin{minipage}{\spl}
      \centering
      $d\bar d\to b\bar b\mu\mu$
    \end{minipage}
               & 4
    & $0.0323(4)$
    & $0.0309(3)$
    & $0.0886(9)$
    & $0.0881(9)$\\
    \hline
    \begin{minipage}{\spl}
      \centering
      $s\bar s\to b\bar b\mu\mu$
    \end{minipage}
               & 4
    & $0.0036(2)$
    & $0.0039(1)$
    & $0.0103(1)$
    & $0.0106(1)$\\
    \hline
    \begin{minipage}{\spl}
      \centering
      $c\bar c\to b\bar b\mu\mu$
    \end{minipage}
               & 4
    & $0.00160(5)$
    & $0.00165(1)$
    & $0.0044(1)$
    & $0.0041(2)$\\
    \hline
    \hline
    All &
    & $76.4(4)$
    & $76.6(4)$
    & $241(1)$
    & $247(1)$\\
    \hline
  \end{tabular}
\end{table}

The search for the light pseudoscalar boson, produced in association
with two $b$ jets and decaying into two muons, was performed at
$\sqrt{s}=8~\text{TeV}$ in the previous CMS
paper~\cite{1707.07283}. It was found that
$\sigma\left(pp\to b\bar b P\right)\times {\rm
  Br}\left(P\to\mu\mu\right)>350~\text{fb}$ is excluded at 95\%
confidence level for $M_{P}=30~\text{GeV}$. To compare the observed
excess with this result we are going to separate the processes with
two $b$ jets in the final state and find the total cross section which
corresponds to the observed fiducial one. In order to do that we have
to find the cut efficiency for the subprocesses with two $b$ jets in
the final state, i.e. we need the total cross sections for these
subprocesses. The CalcHEP results for these cross sections are
summarized in the Table~\ref{tab:bXjet}.

\begin{table}[t]
  \centering
  \caption{ Cross sections for the $pp \to bb X + \dots$ reaction and
    its subprocesses for $Y_{X b \bar b} = 0.01$ and $Y_{X\mu\mu}=1$
    at $\sqrt{s} = 8$~TeV. The errors correspond to integration errors
    reported by CalcHEP.}
  \label{tab:bXjet}
  \begin{tabular}{|c|c|l|l|l|l|}
    \hline
    Subprocess & Mult.
    & $\sigma$ [pb], $S$
    & $\sigma$ [pb], $P$
    & $\sigma$ [pb], $V$
    & $\sigma$ [pb], $A$\\
    \hline
    \begin{minipage}{\spl}
      \centering
      $bb\to bb\mu\mu$\\
      $\bar b\bar b\to \bar b\bar b\mu\mu$
    \end{minipage}
               & 1
    & $0.024(2)$
    & $0.025(1)$
    & $0.048(3)$
    & $0.061(2)$\\
    \hline
    $gg\to b\bar b\mu\mu$
               & 1
    & $1.66(3)$
    & $1.96(3)$
    & $5.68(9)$
    & $5.57(3)$\\
    \hline    
    \begin{minipage}{\spl}
      \centering
      $b\bar b\to b\bar b\mu\mu$
    \end{minipage}
               & 2
    & $0.034(3)$
    & $0.029(1)$
    & $0.072(1)$
    & $0.056(2)$\\
    \hline
    \begin{minipage}{\spl}
      \centering
      $u\bar u\to b\bar b\mu\mu$
    \end{minipage}
               & 2
    & $0.00109(1)$
    & $0.00091(1)$
    & $0.00250(1)$
    & $0.00279(1)$\\
    \hline
    \begin{minipage}{\spl}
      \centering
      $d\bar d\to b\bar b\mu\mu$
    \end{minipage}
               & 2
    & $0.00077(1)$
    & $0.000640(1)$
    & $0.001735(3)$
    & $0.001957(6)$\\
    \hline
    \begin{minipage}{\spl}
      \centering
      $s\bar s\to b\bar b\mu\mu$
    \end{minipage}
               & 2
    & $0.000267(1)$
    & $0.000217(2)$
    & $0.000554(1)$
    & $0.000639(1)$\\
    \hline
    \begin{minipage}{\spl}
      \centering
      $c\bar c\to b\bar b\mu\mu$
    \end{minipage}
               & 2
    & $0.000133(1)$
    & $0.000107(1)$
    & $0.000270(1)$
    & $0.000315(1)$\\
    \hline
    \hline
    All with 2$b$ jets
               &
    & $1.78(3)$
    & $2.07(2)$
    & $5.93(9)$
    & $5.82(3)$\\
    \hline
  \end{tabular}
\end{table}

With the help of the data from Table~\ref{t:pp->bXjet_cut} we can find
the contribution of each subprocess into the reported fiducial cross
section~(\ref{eq:fid.1}) without knowing the coupling constants
$Y_{Xbb}$ and $Y_{X\mu\mu}$:
\begin{equation}
  \label{eq:part.fid.1}
  \sigma^{\rm fid}\left({\rm subprocess}\right)
  = \frac{\left.\sigma^{\rm fid}\left({\rm
          subprocess}\right)\right|_{Y_{Xbb}=10^{-2},Y_{X\mu\mu}=1}}
  {\left.\sigma^{\rm
        fid}\left(\text{All}\right)\right|_{Y_{Xbb}=10^{-2},Y_{X\mu\mu}=1}}
  \times\sigma_{\rm fid.1}.
\end{equation}

Signal selection efficiency $\varepsilon$ depends on the
subprocess. We will calculate it using data from
Tables~\ref{t:pp->bXjet_cut}~and~\ref{tab:bXjet}:
\begin{equation}
  \label{eq:efficiency}
  \varepsilon\left({\rm subprocess}\right) =
  \frac{\left.\sigma^{\rm fid}\left({\rm
          subprocess}\right)\right|_{Y_{Xbb}=10^{-2},Y_{X\mu\mu}=1}}
  {\left.\sigma\left({\rm
          subprocess}\right)\right|_{Y_{Xbb}=10^{-2},Y_{X\mu\mu}=1}}.
\end{equation}

Then we obtain the cross section for individual subprocesses:
\begin{equation}
  \label{eq:sigma_uncut}
  \sigma\left({\rm subprocess}\right)=
  \frac{\sigma^{\rm fid}\left({\rm subprocess}\right)}
  {\varepsilon\left({\rm subprocess}\right)}=
  \frac{\left.\sigma\left({\rm
          subprocess}\right)\right|_{Y_{Xbb}=10^{-2},Y_{X\mu\mu}=1}}
  {\left.\sigma^{\rm
        fid}\left(\text{All}\right)\right|_{Y_{Xbb}=10^{-2},Y_{X\mu\mu}=1}}
  \times\sigma_{\rm fid.1}.
\end{equation}

Then for cross section of subprocesses with two $b$ jets in final
state we get
\begin{align}
  \label{eq:2b_total}
  \sigma\left(pp\to X+2b\text{-jets}\right)\times{\rm
  Br}\left(X\to\mu\mu\right)
  &= \sum\limits_{\text{subprocesses with 2$b$ jets}}\sigma\left({\rm
    subprocess}\right)=\\ \nonumber
  &=
  \frac{\left.\sigma\left(\text{All with 2$b$ jets}\right)\right|_{Y_{Xbb}=10^{-2},Y_{X\mu\mu}=1}}
  {\left.\sigma^{\rm
  fid}\left(\text{All}\right)\right|_{Y_{Xbb}=10^{-2},Y_{X\mu\mu}=1}}
  \times\sigma_{\rm fid.1}=\\ \nonumber
  &=\frac{2.07~\text{pb}}{76.6\cdot10^{-5}~\text{pb}}\times4.1~\text{fb}\approx
  11~\text{pb},
\end{align}
where in the last line we substituted the values for the
pseudoscalar. Let us note that according to A.N.~Nikitenko (private
communication) the cut efficiency for the whole first event category
in case of pseudoscalar is approximately $2.7\cdot10^{-4}$, so the
total cross section is about $15$ pb.

Substituting the data from
Tables~\ref{t:pp->bXjet_cut}~and~\ref{tab:bXjet}
into~(\ref{eq:2b_total}) we get
$\sigma\left(pp\to X+2b\text{-jets}\right)\times{\rm
  Br}\left(X\to\mu\mu\right)$ much larger than the bound at the level
of 350 fb observed in the previous CMS
paper~\cite{1707.07283}. Therefore, the mechanism discussed in this
section cannot be responsible for $X$ production at LHC for any of
$S$, $P$, $V$, $A$.

In the $2HDM$ discussed in Section~\ref{sec:g-2} pseudoscalar $P$ is
produced mainly by radiation from $b$ quark, just like it is described
in this section. Therefore, this model cannot explain experimental
data.

\section{Conclusions}

An extra scalar or vector can describe the resonance discovered
in~\cite{cms}, and simultaneously resolve the disagreement between the
SM prediction for the muon anomalous magnetic moment and its measured
value.

Though $X$ was found in association with at least one $b$ jet, the
simplest model of its production via radiating from $b$ quark line
contradicts the previous CMS paper~\cite{1707.07283}: while the cuts
in the new paper are much stronger (mostly cuts on muons) the fiducial
cross section is at the level of the upper limit on fiducial cross
section from previous paper. To resolve this contradiction, stronger
cuts on muons transverse momentum should not significantly diminish
the number of events, i.e. $X$ should be produced with high transverse
momentum. This can be achieved if $X$ is produced in decays of some
heavy particle, for example, vector-like $B$ quark via
$\bar B_{L}b_{R}X$ interaction term.

Since the New Physics responsible for the observed resonance is
coupled to $b$ quarks in some way, it also can be responsible for the
deviations from SM predictions observed in $B$ decays.

If the existence of $X$ will be confirmed by future experimental data,
it will be a strong additional argument in favor of muon collider
construction.

We are grateful to V.~B.~Gavrilov, who has brought the CMS discovery
to our attention, and to A.N. Nikitenko for valuable comments. We
gratefully acknowledge discussions with R.~B.~Nevzorov. The authors
are supported by RFBR under the grant No.~16-02-00342. S.~I.~Godunov
is also supported by RFBR under the grant No.~16-32-60115.

\bibliographystyle{apsrev4-1}
\bibliography{references.bib}

\end{document}